\documentclass{aa}

\usepackage{graphicx}
\usepackage{txfonts}
\begin{document}

   \title{The Single-mode Complex Amplitude Refinement (SCAR) coronagraph}

   \subtitle{II. Lab verification, and toward the characterization of Proxima b}

   \author{S. Y. Haffert \inst{1}
          \and E. H. Por \inst{1}
          \and C. U. Keller \inst{1}
          \and M. A. Kenworthy \inst{1}
          \and D. S. Doelman \inst{1}
          \and F. Snik \inst{1}
          \and M. J. Escuti \inst{2}
        }

   \institute{Leiden Observatory, Leiden University, PO Box 9513, Niels Bohrweg 2, 2300 RA Leiden, The Netherlands\\
              \email{haffert@strw.leidenuniv.nl}
              \and
              Department of Electrical and Computer Engineering, North Carolina State University, Raleigh, North Carolina 27695, USA}

   \date{Received ? 2017; accepted ? 2017}
 
  \abstract{We present the monochromatic lab verification of the newly developed SCAR coronagraph that combines a phase plate (PP) in the pupil with a microlens-fed single-mode fiber array in the focal plane. The two SCAR designs that have been measured, create respectively a 360 degree and 180 degree dark region from $0.8-2.4 \lambda/D$ around the star. The 360 SCAR has been designed for a clear aperture and the 180 SCAR has been designed for a realistic aperture with central obscuration and spiders. The 360 SCAR creates a measured stellar null of $2-3 \times 10^{-4}$, and the 180 SCAR reaches a null of $1 \times 10^{-4}$. Their monochromatic contrast is maintained within a range of $\pm 0.16 \lambda/D$ peak-to-valley tip-tilt, which shows the robustness against tip-tilt errors. The small inner working angle and tip-tilt stability makes the SCAR coronagraph a very promising technique for an upgrade of current high-contrast instruments to characterize and detect exoplanets in the solar neighborhood.}


   \maketitle
%

\section{Introduction}
We are currently at a breakthrough moment where more and more Earth-like exoplanets are being discovered. Every detection of Earth-sized planets brings us closer to finding life on another planet. The recent discovery of Proxima Centauri b \citep{anglada2016terrestrial} confirms that the solar neighborhood has many planets waiting to be discovered. From current surveys it is also clear that most of the planets in the habitable zone will have a separation close to the diffraction limit of current and future large telescopes. Characterization and detection of these planets can be done through high-contrast imaging, which overcomes the huge contrast between planet and star.

While the indirect methods have been very successful in discovering planets, direct imaging of exoplanets is lagging behind in the number of planets. This is mainly due to the difficulties involved in direct imaging. The largest problem is the close angular position of the planet to the star which is at best a few $\lambda /D$ according to the current statistics of exoplanet orbits \citep{galicher2016survey}. Here $\lambda$ is the wavelength used to observe the system and $D$ the telescope diameter.

The influence of the photon noise on the planet can be reduced by spatially separating the planet signal from the stellar signal. Current and future large optical telescopes have the resolution to resolve planets from their host star. This is done by using extreme adaptive optics (XAO) systems to enable imaging at the diffraction limit on ground based telescopes. The spatial separation also enables the use of coronagraphs to suppress the stellar light. This is the common approach on high contrast imaging (HCI) instruments like SPHERE \citep{beuzit2008sphere}, GPI \citep{macintosh2014first} and SCExAO \citep{jovanovic2015subaru}. 

Combining HCI with high-resolution spectroscopy (HRS) over a broad wavelength range gains further orders of magnitude in contrast close to the star \citep{sparks2002imaging,riaud2007drive}, because high-resolution spectra are able to exploit the difference in spectral lines between the star and planet. This difference can be due to a different Doppler velocity for reflected light and/or due to the presence of different molecular species. This technique has been successfully applied to characterize the atmosphere of several giant exoplanets \citep{brogi2012signature,konopacky2013hr8799c,snellen2014fast}. Recent papers \citep{kawahara2014spectroscopic,snellen2015combining, wang2017hdc} show that this can be used as a robust post processing technique to remove residual stellar speckles which limit current HCI instruments \citep{aime2004usefulness,martinez2012speckles}. \citeauthor{snellen2015combining} simulated a hypothetical Earth-like twin around Proxima Centauri for the European Extremely Large Telescope(E-ELT). The combination of HCI and HRS was able to detect and even characterize the Earth twin. The discovery of an actual Earth-like planet around Proxima Centauri makes this technique even more relevant as \citeauthor{lovis2017} show that an upgraded SPHERE (SPHERE+) can be used to characterize Proxima b if it is coupled to a high-resolution spectrograph. In this approach the focal plane of SPHERE+ would be coupled through a fiber link to the high-resolution spectrograph.

\cite{mawet2017observing} argue that using a single-mode fiber (SMF) link between HCI and HRS instruments has an advantage over multi-mode fibers (MMF). A single-mode fiber is more robust against speckle noise due to the mode filtering capabilities. This property has been appreciated by the interferometry community, where single-mode fibers or waveguides are used to combine and filter multiple beams. \citeauthor{mawet2017observing} considers a system where the coronagraph and fiber injection unit (FIU) act separately on the stellar light. In the companion paper(Paper I) \cite{por2017csmfa} we demonstrate the concept of the SCAR coronagraph, where am pupil plane phase plate is designed that uses the properties of the single-mode fiber to reach a deep null close to the star. This system is related to the Apodizing Phase Plate (APP) \citep{codona2006high,otten2017sky}, which also uses an pupil plane phase optics to create dark holes in the PSF. The main difference is that the APP creates a dark hole by reducing the intensity, while SCAR changes the electric field such that the light can be rejected by a single-mode fiber. The SCAR coronagraph can work over a broad spectral bandwidth with high throughput and is tip-tilt insensitive to a large extent. It is well suited to be used as the interface between high-contrast imaging instruments and high-resolution spectrographs. This is the first system that combines the FIU and coronagraph in a single unit.

The SCAR coronagraph works on the basis of electric field filtering by electric field sensitive photonics. The implementation of this work uses single-mode fibers. With the advances of AO and especially extreme AO it is possible to achieve high coupling efficiency into SMFs \citep{jovanovic2017injection,bechter2016LBTinjection}. The amount of light that couples into a single-mode fiber is defined by the coupling efficiency 
\begin{equation}
\eta = \frac{\left|\int E_{\mathrm{in}}^* E_{\mathrm{SMF}} dA\right|^2}{\int |E_{\mathrm{in}}|^2 dA \int |E_{\mathrm{SMF}}|^2dA}.
\end{equation}
Here $\eta$ is the relative amount of light from $E_{\mathrm{in}}$ that is coupled into a single-mode fiber with mode profile $E_{\mathrm{SMF}}$. For a SMF the mode profile is Gaussian. The light that couples into a SMF is effectively averaged by a Gaussian weighing function. Due to this property if the electric field is zero on average it will not couple into the fiber and is rejected. A remarkable property here is that the intensity does not have to be zero, while the electric field can be. This is a much less stringent requirement than the zero intensity for normal high-contrast imaging. The phase plate is used to modify the point spread function (PSF) in such a way that the stellar light couples very badly into the fiber, but the planet still couples well. This coherent imaging approach is very reminiscent of interferometry where these kind of approaches have been used \citep{angel1986nulling, labadie2007mirnulling}. Because of the benefits of coherent imaging such approaches are now also starting to be exploited for conventional direct imaging \citep{mawet2017observing}.

In this paper we show the first lab verification of the SCAR coronagraph and look at the manufacturing feasibility with Monte Carlo simulations. Section 2 describes the optical setup for the measurements and the lab results. And in Section 3 the manufacturing requirements are derived and a Monte Carlo simulation is done to estimate the expected performance. Section 4 summarizes the results of our study.

\section{Optical setup details and first results}
\begin{figure*}
        \includegraphics[width=\textwidth]{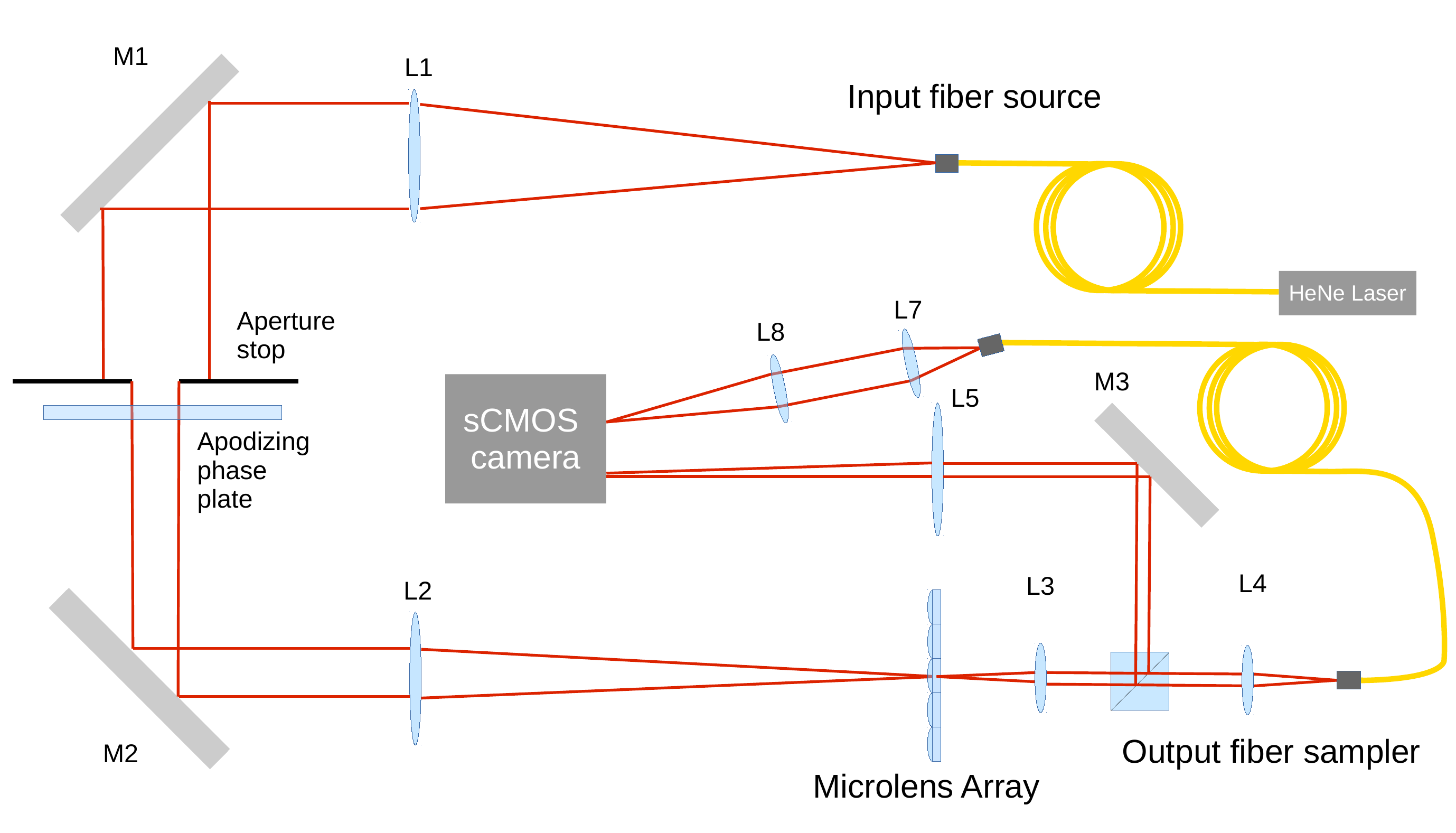}
    \caption{ Schematic of the lab setup. The gray colored surfaces labeled M1 to M3 are mirrors. The blue shaded surfaces L1 to L8 are lenses. The yellow curves represent the single-mode fibers. The central wavelength of the setup is 0.633 nm. }
    \label{fig:labsetup_sketch}
\end{figure*}
\subsection{Lab setup description}
The SCAR coronagraph uses a single-mode fiber array fed by a microlens array. A detailed theoretical description can be found in Paper I. To measure the SCAR performance we created a setup that can emulate the measurements of a microlens fed fiber array. The setup was built in Leiden on a vibration damped optical table in air without any active controlled components. The lab setup used for the measurements can be seen in Figure \ref{fig:labsetup_sketch}. The input source to our setup was a single-mode fiber with a 2.8-4.1 $\mu$m Mode Field Diameter(MFD) at 488 nm, that is fed by a helium neon laser. The fiber was mounted on a XYZ-translation stage. A Thorlabs AC508-1000-B achromatic doublet with a focal length of 1000 mm collimates the fiber. And just before the conjugated plane we placed a pupil stop with a diameter of 3.8 mm. This pupil diameter combined with the 1000 mm focal length ensures that the input source's $\mathrm{MFD}\approx0.02 \lambda/D$ which is much smaller than $\lambda/D$ and therefore creates a good clean point source. 

In the conjugated plane slightly behind the pupil stop we placed the phase plate which modifies the PSF. Another Thorlabs AC508-1000-B achromatic double focusses the light. The resulting PSF is sampled by Okotech APH-Q-P250-F2 hexagonal Micro Lens Array (MLA) with a pitch and diameter of 250 $\mu$m and a 2.18 mm focal length. The scale of the PSF is 166 $\mu$m per $\lambda/D$, this means that each microlens samples 1.5 $\lambda/D$. The MLA sampled PSF is relayed by a set of achromatic doublet lenses which have focal lengths of respectively 100 mm and 75 mm. In the intermediate collimated beam we placed a 90/10 beam splitter which splits 90 percent of the light toward a single-mode fiber and ten percent toward a camera. The transmitted beam was sampled by a LMA-8 photonic crystal fiber of NKT Photonics. This fiber has a mode field diameter of 8.4 $\mu$m $\pm$ 1.0 $\mu$m that is constant as a function of wavelength. The output of this fiber is reimaged on a Andor Zyla sCMOS camera by two positive achromatic doublets. The reflected beam is focused by a 500 mm lens to create an image of the MLA spots onto the same Andor camera. Both the microlens spots and the fiber coupling can be monitored at the same time in this way.
 
This setup was only able to measure the light coupling through the on-axis central lenslet because there is only a single fiber. So to emulate the measurements through an off-axis lenslet the input source was shifted. This relaxes the alignment requirements because the fiber alignment is critical. The fiber has to be aligned within 1/8th of the MFD to create a deep null. Moving the fiber every time to a new lenslet to sample its throughput would be very time consuming due to this alignment requirement. Shifting the PSF is easier because $1 \lambda/D$ is 166 $\mu$m. With a set of digitial micrometer actuators from Thorlabs we were able to shift the PSF with micron accuracy and precision. With this scanning strategy we were able to perform repeatable sub-$\lambda/D$ shifts and measure the throughput as function of shift with respect to the central microlens. The actuator range is two inch which allowed us to scan a range of $\pm 150 \lambda/D$.

\subsection{Fiber alignment procedure}
A misalignment of the output fiber can lead to a reduction in throughput and off-axis nulling. A good alignment is therefore critical. The fiber alignment is done in multiple steps to ensure good nulling. First the input source is aligned on the central microlens without a phase plate in the beam. The microlens spots, which can be viewed on the camera, should be radially symmetric in intensity after this first step because the PSF is radially symmetric. During the second step the fiber is coarsely aligned to find the brightest spot in the field which is done by moving the fiber such that the brightest spot can be seen by eye on the fiber face. The last step takes care of the fine alignment by moving the fiber such that the fiber throughput is maximized.

\subsection{Apodizing phase plate designs}
\begin{figure}
        \includegraphics[width=\columnwidth]{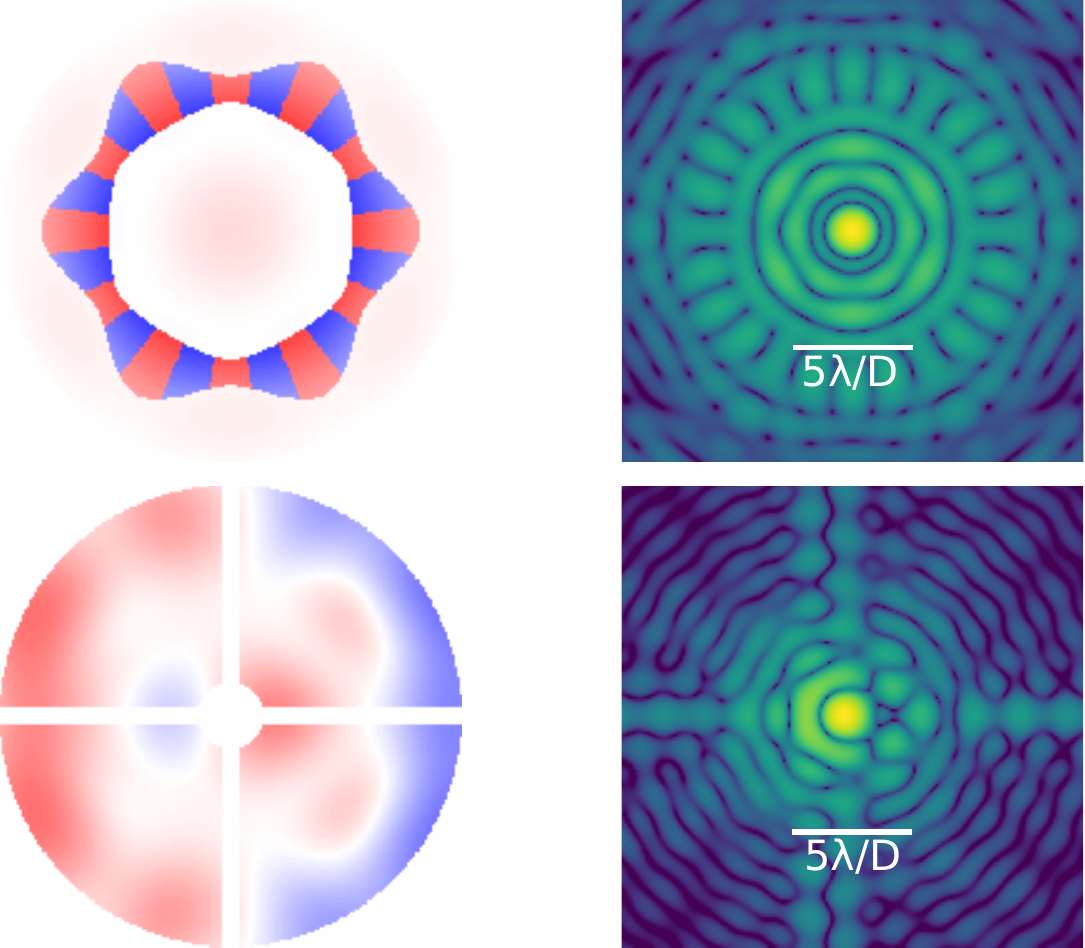}
    \caption{Left column shows theoretical phase pattern. Here blue is $\pi$ phase, white is zero phase and red is $-\pi$ phase. The corresponding PSF is shown on the right with a 5 $\lambda/D$ scale bar. Both images of the PSF are on log scale with the color scale shown on the right. The first phase pattern creates a 360 dark region for a clear aperture. The second pattern creates a one-sided dark region for a pupil that has a central obscuration and spiders.}
    \label{fig:app_patterns}
\end{figure}

Two different SCAR designs have been measured in our setup. The first design is a phase pattern that generates a 360 degree dark region in the first ring of lenslets around the PSF with a contrast of $5\times10^{-5}$ and a throughput of 30 percent. The throughput includes the coupling efficiency into the SMF. The second design is a 180 degree phase pattern with a central obscuration and spiders. This creates a one-sided dark region next to the PSF with a contrast of $1\times10^{-5}$ and a 60 percent throughput. Both patterns are designed for a spectral bandwidth of 20 percent. The phase patterns with their corresponding PSFs can be seen in Figure \ref{fig:app_patterns}.

\subsection{Liquid crystal plate}
\begin{figure}
        \includegraphics[width=\columnwidth]{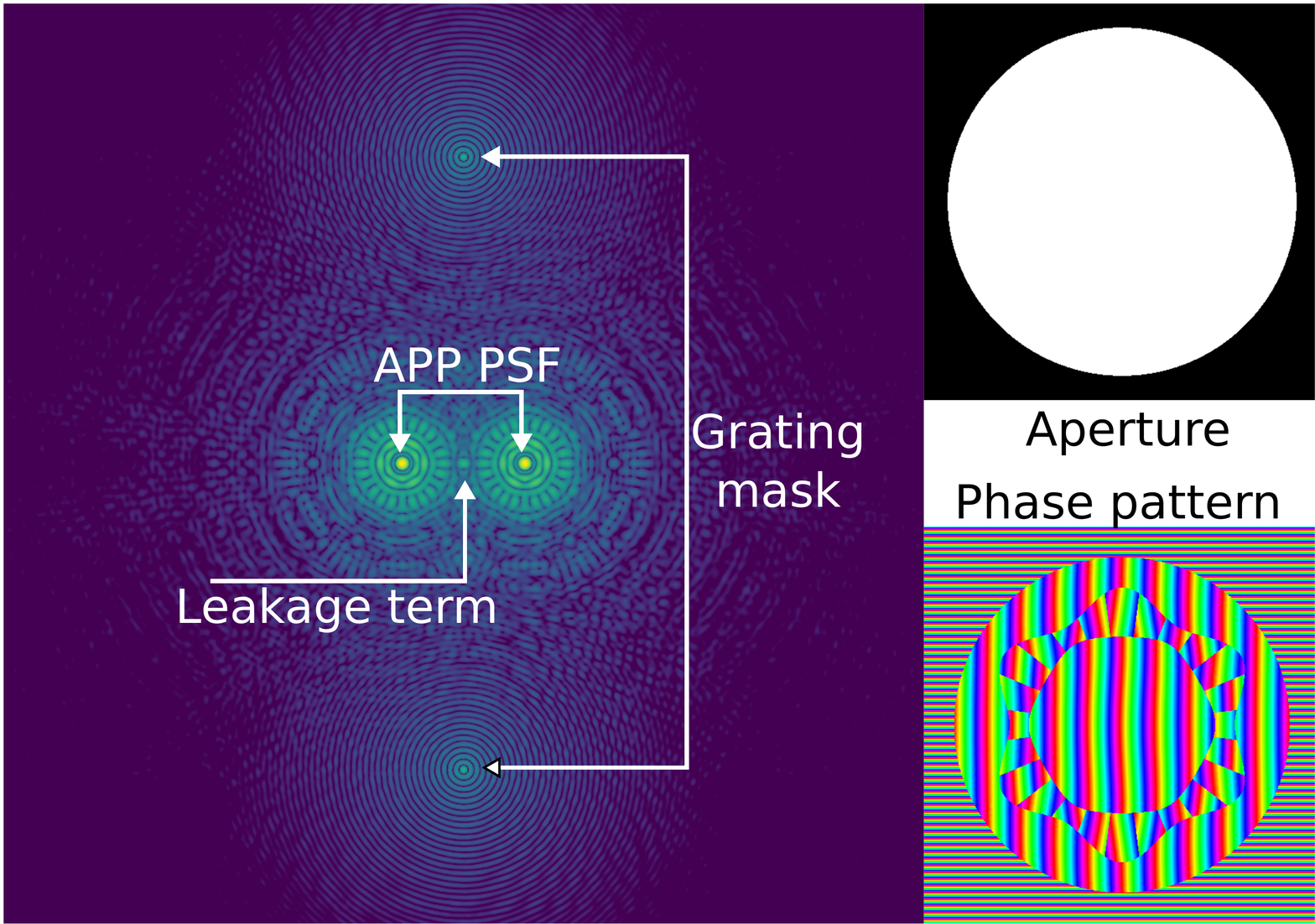}
    \caption{Left: Simulated focal plane of the phase pattern. Right: Phase and aperture. On the focal plane several PSF's can be seen. The left and right PSFs are the PSFs that contain the phase pattern. The top and bottom correspond to the light that scatters away due to the grating mask. And in the center there is a 1 percent leakage term. The horizontal PSFs are at $\pm 10 \lambda/D$ and the aperture PSFs are at $ \pm 50 \lambda/D$ }
    \label{fig:app_grating}
\end{figure}

The phase plates are manufactured by using a direct write approach where the fast axis of the liquid crystals is written with a laser \citep{snik2012vapp, miskiewicz2014directwrite}. The liquid crystals add a geometric phase to the incoming light which only depends on the angle of the fast axis. This then acts as an achromatic phase pattern with a chromatic piston term which can be ignored. An important aspect is that it acts on circular polarized light. Left circular and right circular polarizer light both get the same phase pattern but with a change of sign. Because of this it is important to separate the two polarizations. The separation is done by adding a tilt to the phase pattern, and because both polarization get opposite phase they split into different directions. This is also done in the grating-vAPP \citep{otten2014gratingAPP}. If the liquid crystal plate is not perfectly half-wave there will be a leakage term that does not see the phase pattern. The leakage creates a normal Airy pattern. This can be seen in Figure \ref{fig:app_grating}.The leakage term can limit the contrast if the retarder substantially deviates from half wave. One way to reduced the effect of the leakage is to separate it like the grating-vAPP or use a technique like the double-grating vAPP where the leakage is scattered away \citep{doelman2017liquidcrystals}. For the results in this paper we used the grating-vAPP approach.
The additional pair of spots in Fig. \ref{fig:app_grating} are created by the grating mask \citep{doelman2017liquidcrystals}. The grating mask is a phase tilt that is applied outside the aperture. All the light that falls outside of the defined aperture diffracts due to the phase tilt. With the grating mask we can create a well defined aperture shape and size. Our phase plates were made with 8.75 $\mu$m LC pixels and the pupil itself is 3.8 mm allowing for 434 pixels across the pupil. The aperture spots fall off quickly enough that they do not influence the modified PSFs.

\subsection{Lab setup results}
\begin{figure*}
        \includegraphics[width=\textwidth]{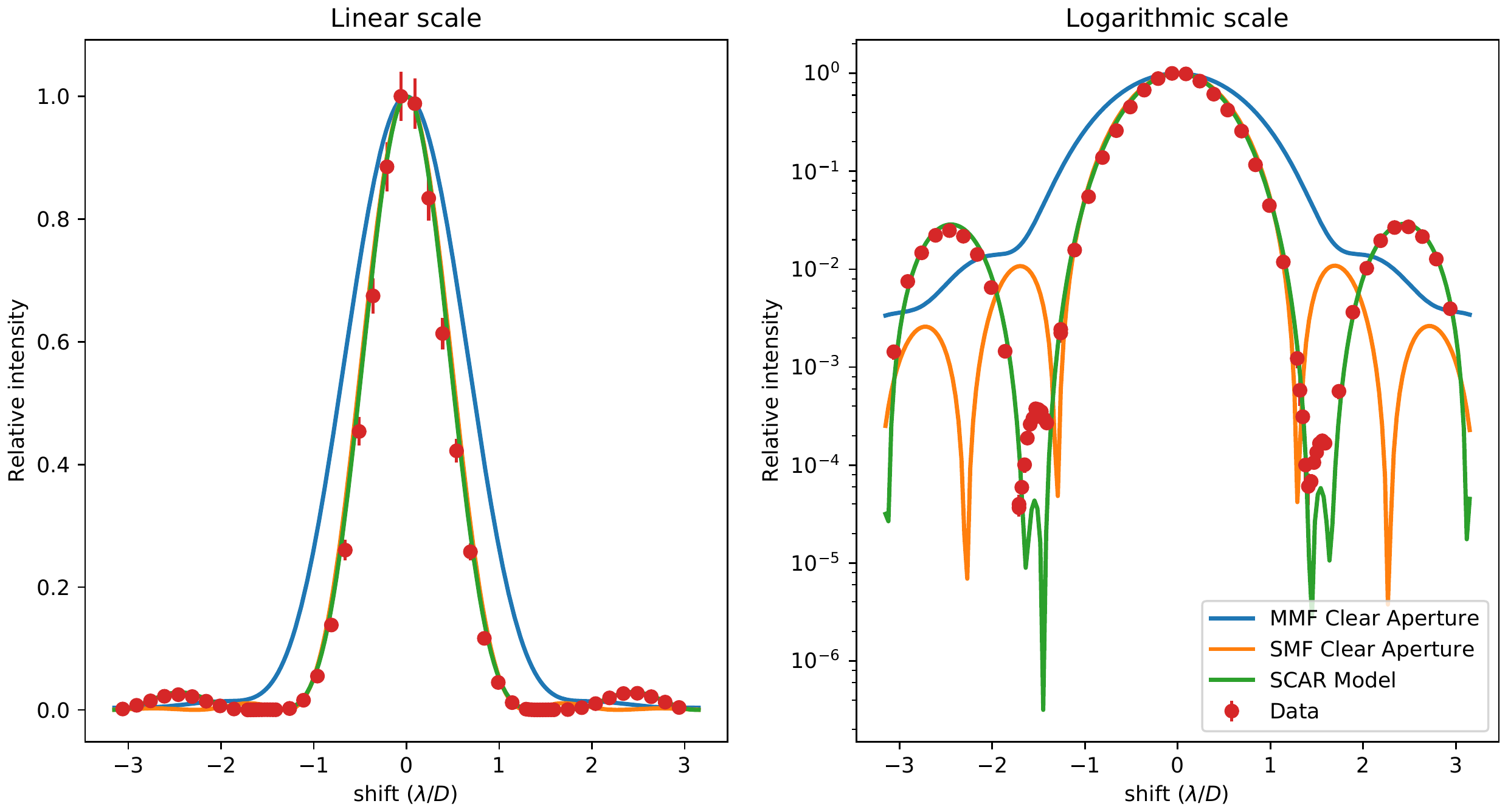}
    \caption{Left: Relative throughput on a linear scale as function of relative shift with respect to the microlens center. The red points are the measurements with errorbars due to random errors. The green line shows the model of this SCAR design. Right: Throughput on a logarithmic scale. The null is uneven between left and right and not as deep as designed. This is suspected to be caused by 10 nm rms residual low-order aberrations. The blue line shows the amount of light that falls on the microlens with a clear aperture and without SCAR. This is comparable to the contrast curve for normal imaging or using a multi-mode fiber. The orange line shows the normalized throughput with a unobstructed aperture and a SMF. The SMF shows a gain compared to the MMF. }
    \label{fig:360_APP_data}
\end{figure*}

\begin{figure}
        \includegraphics[width=\columnwidth]{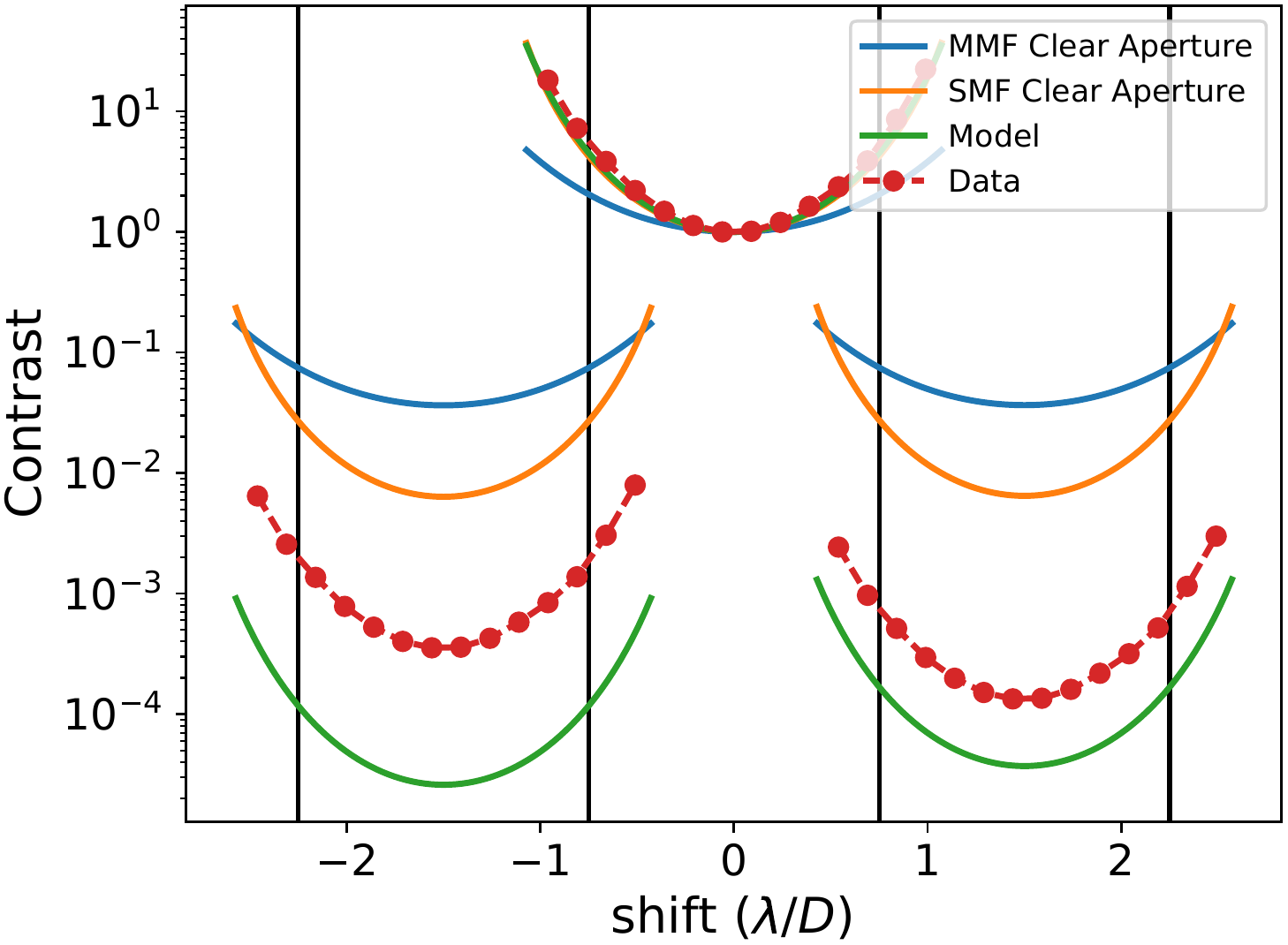}
    \caption{Contrast as a function of position on the microlens array. The black lines show the borders of the microlenses. The blue lines show the contrast for a multi-mode fiber. The orange lines show the contrast for a single-mode fiber. A single-mode fiber already provides extra contrast compared to a multi-mode fiber. The green lines show the model of the SCAR coronagraph, and the red dots are the measurements.}
    \label{fig:360_contrast}
\end{figure}

The throughput as function of PSF position for the 360 SCAR can be seen in Fig. \ref{fig:360_APP_data}. The measured throughput is overlayed on the model curve. Given that the variables of our model are not fitted but taken as is from the manufacturing specifications, the measurements and the model agree very well. As said before this is the relative throughput as function of distance from a microlens center. This is not a contrast curve. For this system with microlenses the contrast curve is discrete and it only changes when going to another microlens as shown in Figure \ref{fig:360_contrast}. The contrast is defined as the ratio between the coupling of the on-axis object and the off-axis object. Figure \ref{fig:360_contrast} shows that the contrast changes as the source moves over the microlens due to a change in throughput.

On the linear scale it is easier to read off the throughput loss as the source is shifted. The 50 \% throughput is at an offset of 0.5 $\lambda/D$. We define the inner working angle as the smallest angular separation where the throughput of the companion is equal to 50\% of the maximum, and the outer working angle as the largest angular separation where the throughput is equal to 50\% of the maximum. A source has a maximum throughput when it is in the middle of a microlens and needs to shift by 0.5 $\lambda/D$ for it to reach half the maximum throughput. Therefore we can define the effective inner and outer working angles as $1.1 \lambda/D$ and $2.1 \lambda/D$ for the first ring of lenslets.

At $0.75 \lambda/D$ the PSF is precisely on the edge between two microlenses. At this position 25\% throughput remains in each of the two fibers. So a total throughput of 50\% is achieved by combining multiple fiber outputs. If a binary system is observed with field rotation then the relative throughput of the source fluctuates between 50\% and 100\% as it rotates over the lenslets. For an off-axis source that is on the edge between the central lenslet and an off-axis lenslet then we can only capture 25\% of the off-axis source, but with an impressive source separation of $0.75 \lambda/D$.

The contrast changes as a source moves over the micro-lens array because the throughput changes. This is shown in Figure \ref{fig:360_contrast}. The MMF shows the intrinsic contrast for normal imaging. Compared to a MMF a SMF already increases the contrast by a factor of 10 due to the rejection of the nonGaussian modes that are in the PSF. The 360 SCAR can in theory reach an average contrast of $5\times10^{-5}$ but we are limited by static aberrations in the system, which limit the contrast to $3\times10^{-4}$ at $-1.5 \lambda/D$ and $2\times10^{-4}$ at $1.5 \lambda/D$. The decrease of contrast can be attributed to 10 nm rms low order wavefront aberration, but this is a very rough estimate. The change in contrast is symmetric around $1.5 \lambda/D$. The contrast of the right fiber stays below $1\times 10^{-3}$ and the left fiber stays below $2\times 10^{-3}$. Figure \ref{fig:360_contrast} also shows that if the source is on a different microlens it can still couple into the other microlenses and reach an acceptable contrast.

The broad gap around $1.5 \lambda/D$ in Figure \ref{fig:360_APP_data} gives insight into the wavelength scaling and the effects of jitter. When the wavelength changes then $\lambda/D$ changes and we have to check the stellar throughput at a different part of the curve. So the width of the gap is a measure for the bandwidth. The spectral bandwidth is then roughly $2 * \Delta \theta / \theta_0 \approx 0.2$ with $\Delta \theta$ the gap width and $\theta_0$ the gap center. Next to the wavelength response it also says something about the monochromatic jitter resistance. If the star jitters a bit then the off-axis throughput is still low because of the gap. The gap where we still achieve the contrast for the 360 SCAR is between $-1.8 \lambda/D$ and $-1.3 \lambda/D$. The gap on the right is from $1.36 \lambda/D$ to $1.7 \lambda/D$. This demonstrates that the coronagraph should be able to handle $\pm 0.15 \lambda/D$ monochromatic tip-tilt residuals. The SMF without any pupil phase optic in contrast has a very narrow rejection area. There is only one position at which it nulls the star and if there is a small amount of jitter the contrast quickly deteriorates to about $10^{-2}$. This shows the advantage of SCAR, which creates a broad dark area. Because the system is completely passive it is also very robust. The dark region remained dark over several weeks.

\begin{figure*}
        \includegraphics[width=\textwidth]{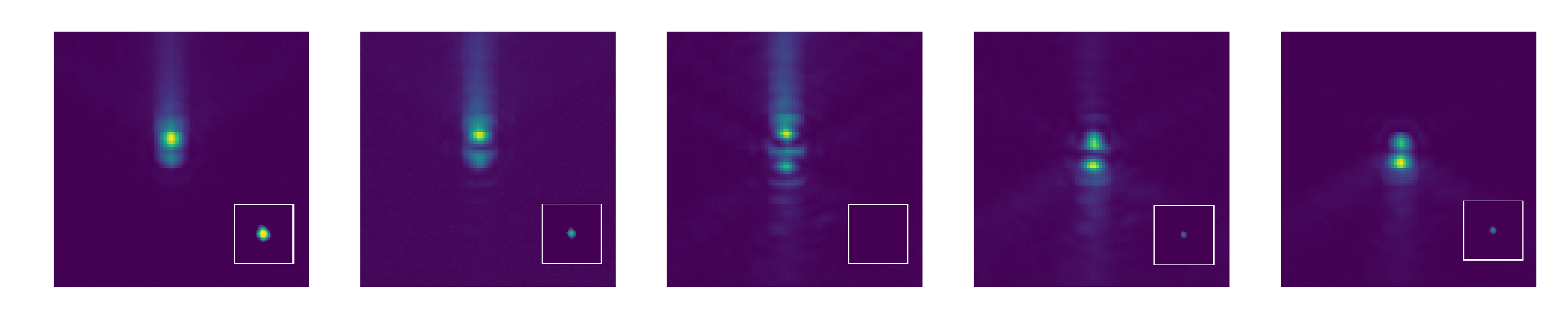}
    \caption{Measured microlens spot structure for the 360 SCAR. The inset shows the throughput of the single mode fiber. The deepest null occurs with a triple spot structure, which is a second order null  due to the two zero crossings. The white circle shows the position and MFD of the fiber. }
    \label{fig:spot_slice}
\end{figure*}

The filtering effect of the -mode fiber can be seen in Figure \ref{fig:spot_slice}, where the measured microlens spots around the deepest null are shown together with the fiber throughput. The spot has a triple peak structure that is created by the phase plate. The triple peak structure is the feature that increases the bandwidth and tip-tilt stability. This is characteristic of a second-order null. The triple peak structure can be seen in the middle frame of Figure \ref{fig:spot_slice}. While the change in total power from frame to frame is small, the fiber throughput changes drastically. This shows the modal filtering capability of single mode fibers.

\begin{figure*}
        \includegraphics[width=\textwidth]{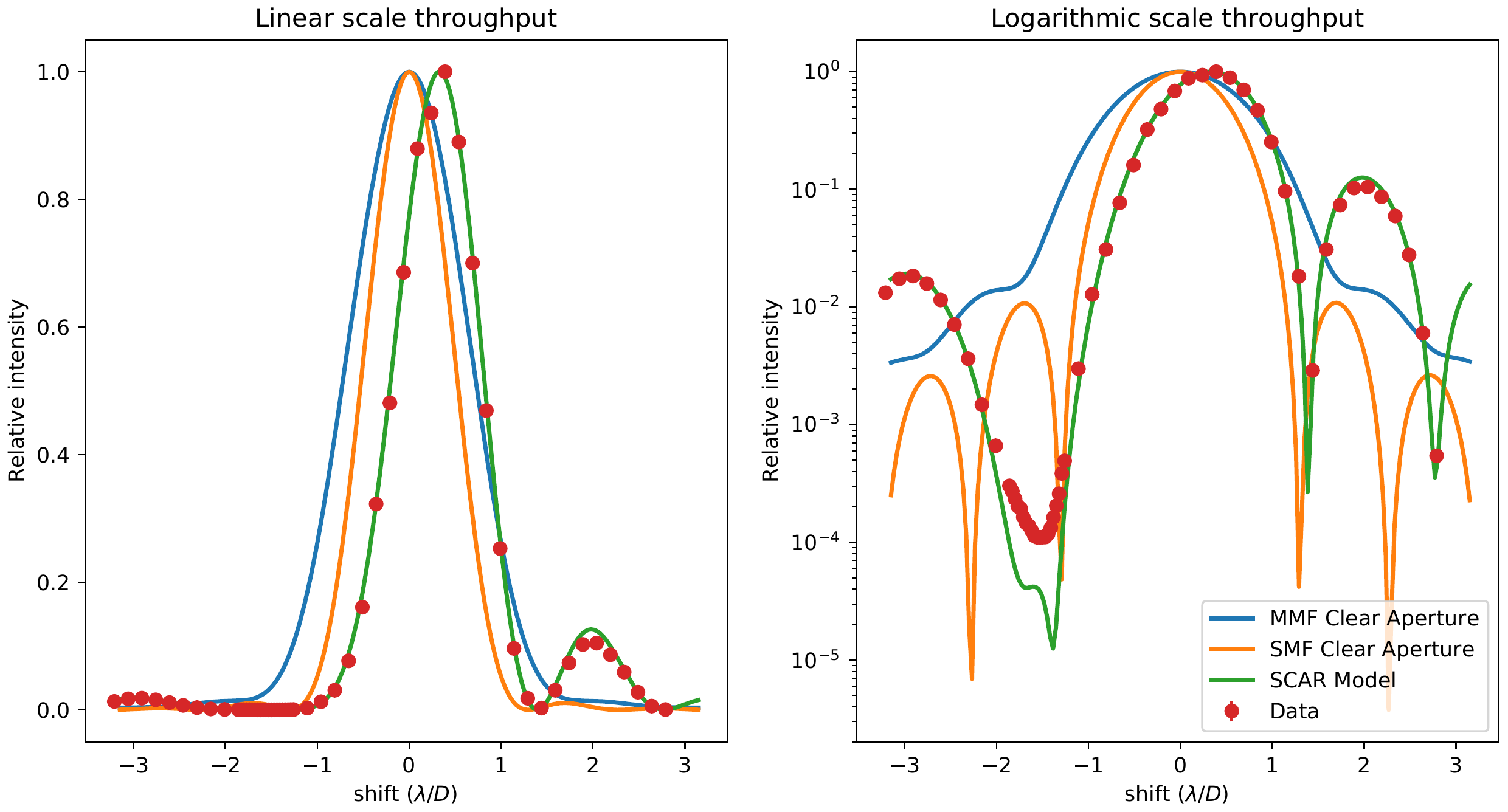}
    \caption{Left: Relative throughput on a linear scale as function of relative shift with respect to the microlens center. The green points are the measurements. Errorbars are included but are smaller than the size of the plotting symbols. The errorbars are very small and show that random errors can not explain the difference between the model and the measurements. The orange line shows the nonfitted model of this SCAR design. Right: Throughput on a logarithmic scale. The deepest contrast that we reach is $1.15\times10^{-4}$. The throughput curve is also flat between -1.6 and -1.4 $\lambda/D$. The measured null does not reach the design null due to residual low order aberrations aberrations on the order of 10 nm rms. The blue curve shows the amount of light that falls in a microlens of a clear aperture, which shows the contrast without SCAR. This is comparable to the raw contrast curve for normal imaging or using a multi-mode fiber. The orange line shows the normalized throughput with an unobstructed aperture and a SMF. The SMF shows a gain compared to the MMF. }
    \label{fig:180_APP_data}
\end{figure*}

\begin{figure}
        \includegraphics[width=\columnwidth]{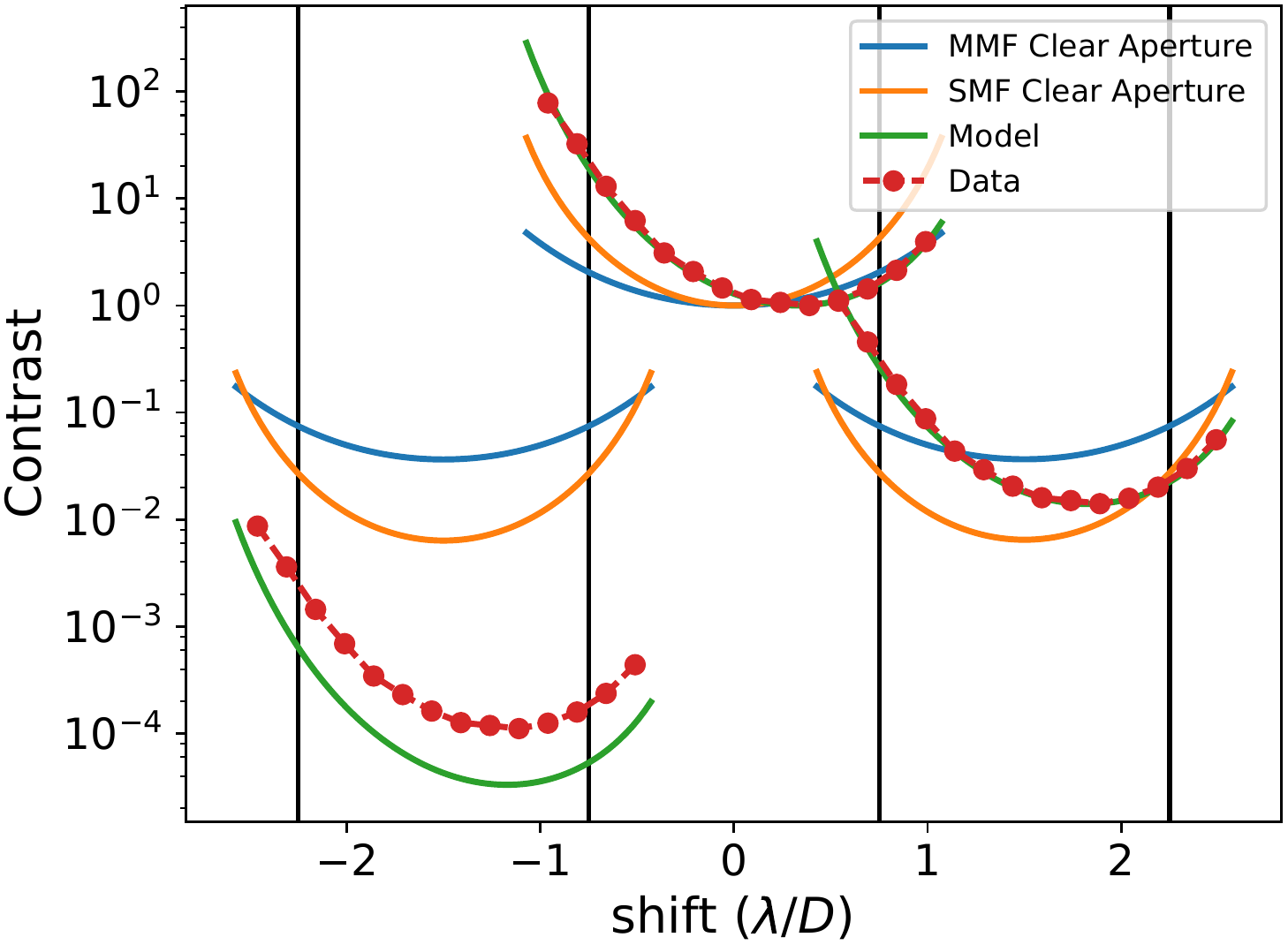}
    \caption{Contrast as a function of position on the microlens array. The black lines show the borders of the microlenses. The blue lines show the contrast for a multi-mode fiber. The orange lines show the contrast for a single-mode fiber. A single-mode fiber already provides extra contrast compared to a multi-mode fiber. The green lines show the model of the SCAR coronagraph, and the red dots are the measurements.}
    \label{fig:180_contrast}
\end{figure}

The throughput results for the 180 SCAR can be seen in Figure \ref{fig:180_APP_data}. The measurements reach a contrast of $1.15\times10^{-4}$. There is a mismatch between the measured and simulated throughput curves, but we can see that compared to the 360 design this design reaches a deeper contrast. The deepest part is also relatively flat between -1.7 and -1.4 $\lambda/D$. Because of the flat response the design can handle tip-tilt errors of $\pm 0.15 \lambda/D$. The corresponding constrast curves are shown in Figure \ref{fig:180_contrast}. The contrast curves are asymmetrical due to the asymmetric PSF that is created by the 180 SCAR.


Apart from maximizing the contrast it is important to optimize the absolute throughput of the planet light through the fiber. For the proposed system we have defined the coupling not as the total amount of light that couples into the fiber, but as the amount of light that falls within the micro-lens aperture that couples into the fiber. The total amount of light that couples into the fiber depends on how much of the PSF the micro-lens captures. Therefore there is a trade-off between spatial resolution and throughput \citep{por2017csmfa}. The 360 SCAR has a maximum coupling of 87\% percent for a single micro-lens. Due to the hexagonal shape of the micro-lens the maximum coupling efficiency is slightly higher than the 82\% theoretical maximum for circular apertures \citep{shaklan1988}. Our lab measurement of the coupling was defined as the ratio between the fiber spot intensity and the MLA spot intensity. After correcting for the uneven beam-splitter between the two spots our absolute coupling is $76\pm3$ \%. This includes Fresnel losses at the interface of the fiber, fiber propagation losses and roughness due to the polishing of the fiber. The measured coupling is very close to the theoretical maximum if we consider these losses. The absolute throughput is the product of the coupling, a correction for the spatial sampling and the Strehl ratio of the phase plate. The measured absolute throughput is 26 $\pm$ 1\%.

\section{ Tolerance simulation analysis }
\cite{lovis2017} show that within certain assumptions about Proxima b that it can be characterized by combining SPHERE+ with a high-resolution spectrograph (ESPRESSO in this case). Several challenges are to be solved. Current coronagraphs on SPHERE are not able to suppress the stellar diffraction halo at the position of Proxima b with the required contrast, and the AO system is not good enough at the angular separation of Proxima b. \cite{lovis2017} propose SPHERE+ where both the coronagraph and the AO system are upgraded. Switching from a Shack-Hartmann wavefront sensor to a Pyramid wavefront sensor would greatly improve performance at small inner working angles \citep{verinaud2005pyramid, fusco2006}. For SPHERE+ \citeauthor{lovis2017} assumed a hypothetical coronagraph. The listed requirements of this coronagraph are:
\begin{itemize}
\item A contrast of at least 5000.
\item A relatively broad wavelength range of at least 20 percent.
\item The stellar rejection region should encompass the orbit of Proxima Centauri b, which has a maximum estimated separation of 36 milliarcseconds (2.2 $\lambda/D$ at $0.7$ $\mu$m).
\item An inner working angle of 1 $\lambda/D$ to reach the full resolving power of the telescope.
\item Either circular or asymmetric dark holes.
\item The coronagraph should be able to handle tip-tilt errors within 3 mas.
\end{itemize}
We designed a phase plate for the SCAR coronagraph that would fit these requirements in Paper I. This SCAR design is able to null a circular area from 0.8 to 2.4 $\lambda / D$, which is large enough as Proxima b has a maximal separation of 2.2 $\lambda/D$ at 0.7 $\mu$m. The design bandwidth is 10\%. Within this bandwidth the raw contrast is $3\times 10^{-5}$, which is 10 times higher than required. Outside of the design bandwidth it still works well as the contrast stays below $1\times 10^{-4}$ up to 20\% bandwidth. The bandwidth of the coronagraph is slightly smaller than required. With this design we would be able to reach almost all requirements of the coronagraph for the characterization of Proxima b. The final requirement is necessary as the tip-tilt jitter of SPHERE is 3 mas \citep{fusco2016SAXO}.

It is important to estimate the effects of manufacturing errors which change the reachable contrast. Our target contrast including manufacturing errors is $10^{-4}$. There are three major parts that can influence the final performance. The SCAR phase plate is manufactured with the same techniques as the APP. Because the APP has demonstrated on sky that it can achieve a high contrast (\cite{otten2017sky}). For the required contrast levels we can safely assume that the manufacturing errors in the phase plate are negligible. The other two aspects are the manufacturing tolerances on the fiber array and the residual wavefront error after the SPHERE AO system. We focussed on the manufacturing tolerances of the fiber injection unit because residual wavefront errors can not be solved by the coronagraph.

\subsection{Fiber alignment tolerance}
\begin{figure*}
        \includegraphics[width=\textwidth]{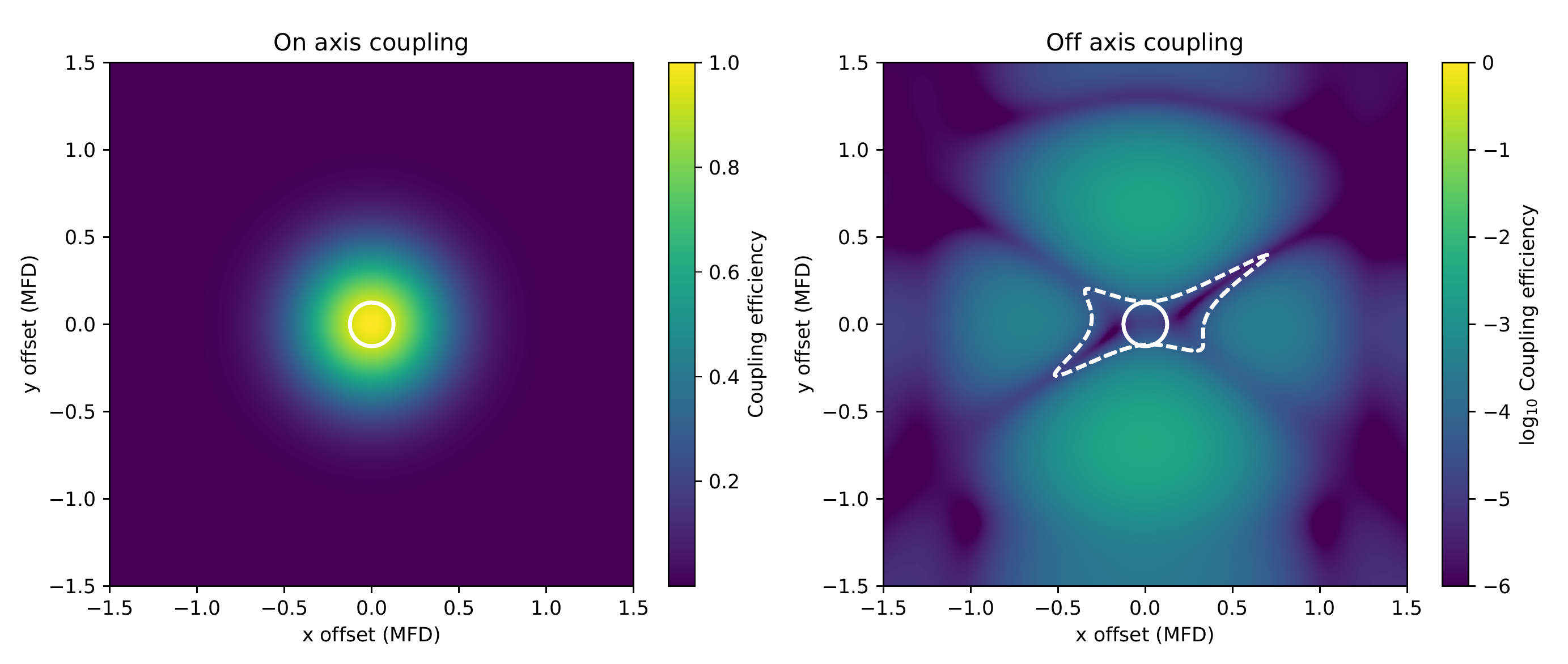}
    \caption{Throughput of an on-axis source as function of fiber misalignment is shown on the left. The right shows the off-axis contrast as function of misalignment. The white contour encircles the area where the contrast is below $1 \times 10^{-4}$. The white circle is the largest circle that fits within the contour with a radius of 1/8th of the mode field diameter} 
   \label{fig:fiber_tolerance}
\end{figure*}

To couple well into a single mode fiber it is necessary to have a good alignment of the fiber with respect to the center of the microlens surface. For normal operations of a single mode fiber, which is getting in as much light as possible, the alignment tolerance is already strict. For the fiber coronagraph the alignment tolerance becomes even stricter. In Figure \ref{fig:fiber_tolerance} the throughput as function of the fiber shift is shown. The figure shows the throughput of the central lenslet ( which is the planet coupling ) and the throughput on an off-axis lenslet ( which is the nulling of the star ) as function of fiber offset. The white circle shows the largest fitting circular region where the contrast is still below $10^{-4}$. In the same region the on axis relative throughput is above 95 percent. This shows that injecting light into a fiber is easier than using a fiber to cancel the light as the off-axis throughput surface is a much steeper function of misalignment. The white circle has a radius of $1/8$ of the mode field diameter of the fiber. The fiber alignment should be within this diameter to reach the required contrast.

Step index single mode fibers are the most used single mode fiber and they have a mode field diameter with a size around 5 $\mu$m at a wavelength of 700 nm. Given this size the alignment tolerance would be roughly $\pm 0.6$ $\mu$m. This is very strict and most manufacturing procedures have an alignment tolerance of 1 $\mu$m. The submicron tolerance can be circumvented with the use of fibers with a larger mode field diameter. Fibers with a large mode field diameter can be made with Photonic Crystal Fibers(PCF). Photonic crystal fibers do not use internal reflection to guide the light as step index fiber do, but use the geometry of the fiber structure. Specific geometric configurations create bandgaps which allows certain optical modes to propagate (\cite{corbet2006PCF}). Large mode area photonic crystal (LMA) fibers are fibers with a large mode field diameters which can be up to 50 $\mu$m (\cite{stutzki2014vlma,jansen2012vlma}) and only allow propagation of the fundamental mode. Another advantage of the PCFs is their endlessly single mode property that allows for a very large wavelength range to be propagated through the fiber, which is convenient for spectrographs. The manufacturing tolerances of 1 $\mu$m would require a fiber with a MFD of at least 8 $\mu$m. Fibers with a mode field diameter of 12.5 $\mu$m are readily available and would fulfill the alignment tolerance requirement. These fibers have a less strict tolerance which is $\approx 1.5\mu$m. With the PCFs the alignment of the fibers should not be an issue, and a manufacturer has been identified.

The LMA fibers have a constant MFD but the numerical aperture (NA) depends on wavelength because of the conservation of etendue. The common step-index fibers have the opposite behaviour: the NA is fairly constant but the MFD changes with wavelength. The microlens launches the light with a constant NA into the fiber, which can lead to a mode-mismatch in the case of LMA fibers. The mode-mismatch reduces the coupling as a function of wavelength and is optimal only for a single wavelength. In Paper I we simulated and designed the phase plates with the LMA fibers in mind. There we saw that the throughput as a function of bandwidth changes very slowly. If the wavelength changes by 50\%, the throughput drops from ~55\% to 40\%. Within the design bandwidth of 20\% the throughput varies between 50-55\%. For our bandwidth this will lead to a slightly lower efficiency compared to step-index fibers.

\subsection{MLA surface}
\begin{figure*}
        \includegraphics[width=\textwidth]{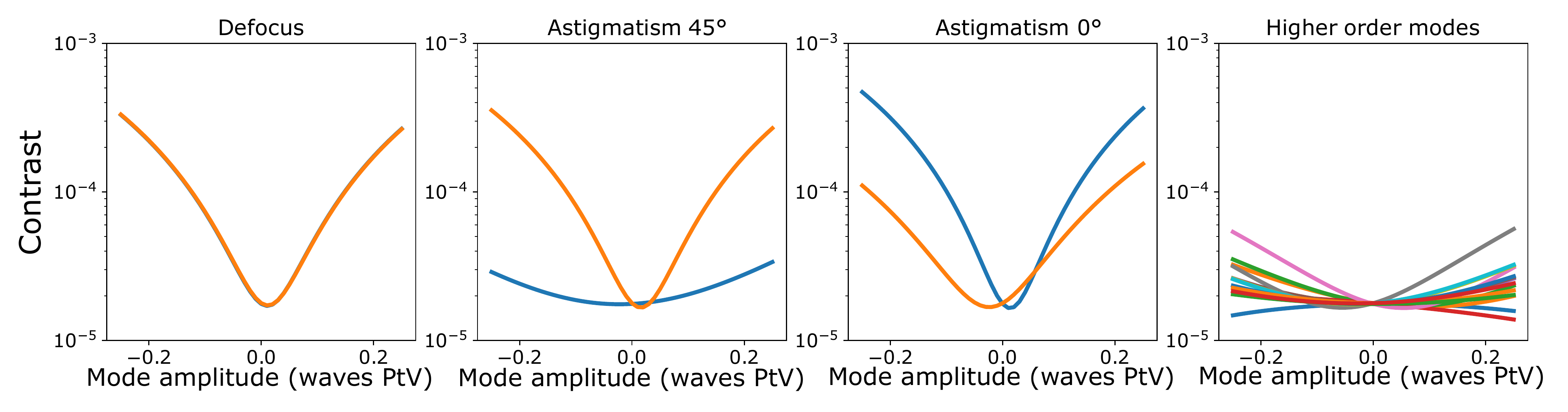}
    \caption{Effect of wavefront errors in the microlenses on the contrast due to a certain Zernike mode. The first three panels show the effects on two different fibers. Due to symmetry the other four fibers behave in the same way as these two. The radial symmetry of the defocus mode causes the same effect in all fibers, therefore the curves of the two fibers overlap. In the rightmost panel the effects of Zernike modes 6 to 10 are shown. Focus and astigmatism are the most dominant wavefront errors for lowering the contrast. }
    \label{fig:zernike_tolerance}
\end{figure*}

The microlens array will have surface errors. Therefore it is important to know how much the errors influence the final contrast. The effects of Zernike wavefront errors on the microlens array are shown in Figure \ref{fig:zernike_tolerance}. For the low order Zernike modes there are two curves per figure which show the contrast for two different fibers. Due to symmetry the other four fibers behave in the same way as one of these two. The most important surface errors are defocus and astigmatism. Both rapidly degrade the contrast. The higher order Zernike modes have almost no influence on the contrast this can be seen in the fourth panel in Figure \ref{fig:zernike_tolerance}. 
The defocus can be partially compensated by moving the fiber in the axial direction. The astigmatism can not be compensated by changing the alignment of the fibers, so it puts a requirement on the surface deviation of the microlens array. The astigmatism of the microlenses should be smaller than $\lambda / 6$ peak to valley to reach the required $10^{-4}$ contrast.

\subsection{Fiber mode shape}
From our investigations we noticed that the mode profile is not very important. In this analysis we changed the mode radius, eccentricity and the orientation of the resulting multivariate Gaussian mode. The mode field radius can change by $\pm$10$\%$ while still create a contrast below $1\times 10^{-4}$. Usual manufacturing constrains on the mode field radius are also within $\pm$10$\%$, therefore the mode field radius is not an issue. Eccentricities up to 0.5 and the orientation of the ellipse had no significant impact on the throughput and contrast. The average contrast went up from $2.7 \times 10^{-5}$ to $\approx 3 \times 10^{-5}$ due to these two parameters. From this we conclude that the tolerances on the mode profile of the fiber will be easily met.

\subsection{FIU Monte Carlo analysis}
\begin{figure}
        \includegraphics[width=\columnwidth]{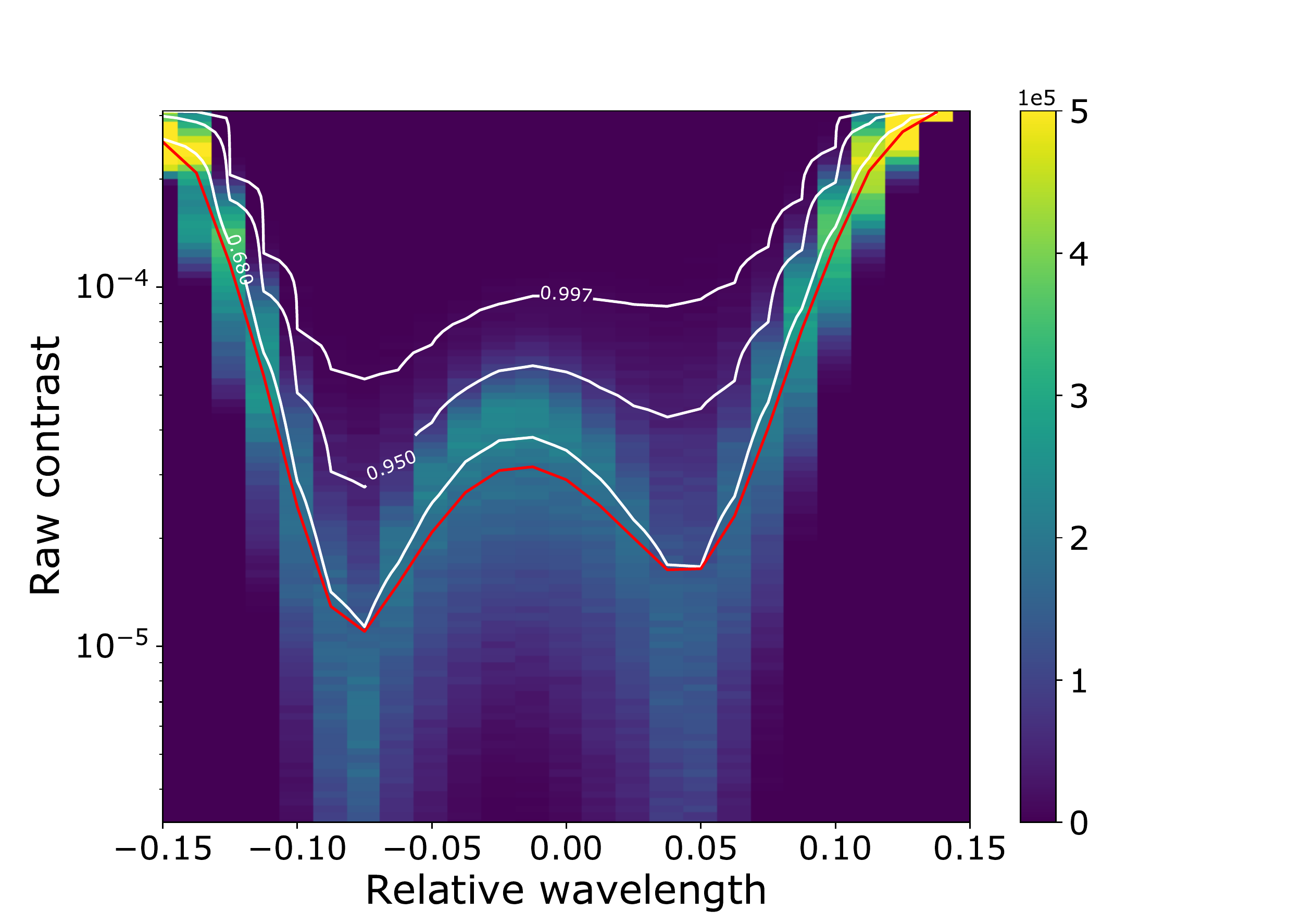}
    \caption{Results of the Monte Carlo tolerance analysis as function of wavelength. The red curve shows the expectation value of the contrast. The white curves are the 0.68, 0.95 and 0.997 percentile confidence limits. The analysis indicates that within specified tolerances a contrast below $1 \times 10^{-4}$ can be reached over a 15 percent bandwidth. Within the full 20 percent bandwidth the contrast is below $2 \times 10^{-4}$.}
    \label{fig:full_tolerance}
\end{figure}

In the previous section several manufacturing errors have been looked at independently of other errors. A Monte Carlo analysis is performed to estimate the degradation of the coronagraph due to all manufacturing errors. The Monte Carlo analysis generated 3000 realizations of the system within the parameter space given in Table \ref{tab:mc_tol}. The results of this analysis can be seen in Figure \ref{fig:full_tolerance}, where the probability density function as function of wavelength and contrast is plotted. The expected performance of the coronagraph plotted on the figure is well under $1\times10^{-4}$. The three sigma threshold shows that within the manufacturing specifications we will reach the required contrast with very high certainty. The expected stellar nulling within 15\% bandwidth is below $3\times 10^{-5}$, within 20\% below $1\times 10^{-4}$ and within 25\% below $2\times 10^{4}$. After correction for throughput variations this meets the required specifications for characterizing Proxima b with SPHERE+.

\begin{table}
  \centering
  \caption{The Monte Carlo parameters for the tolerance analysis. The parameter column shows which parameters are varied and the distribution column shows what distribution is assumed for a parameter.}
  \label{tab:mc_tol}
  \begin{tabular}{l|ll}
    Parameter           & Distribution parameter  & Distribution \\ \hline
    Mode field diameter & $\pm 5$ percent         & Uniform      \\
    Fiber misalignment  & $\pm 1/12$ MFD          & Uniform      \\
    Microlens focus     & $\sigma = 0.01$ percent & Gaussian     \\
    Microlens astigmatism & $\sigma = \lambda/4$    & Gaussian
  \end{tabular}
\end{table}

\section{Conclusions}
Combining a pupil plane phase plate with a single mode fiber array creates a new coronagraph that can be used to search for planets very close to their host star. Adding the high-resolution spectroscopy post-processing makes this a very robust system for exoplanet characterization. From the lab measurements and tolerance simulations we can conclude that:
   \begin{enumerate}
      \item We have shown a proof of principle for the SCAR coronagraph in the lab and reached a contrast $1\times 10^{-4}$ for the 180 SCAR and $2-3\times 10^{-4}$ for the 360 SCAR. These contrasts limit were due to residual wavefront errors, which requires an active system to remove.
      \item The monochromatic tip-tilt stability of the coronagraph has been measured and is estimated to be on the order of $\pm 0.15 \lambda/D$ for both the 360 and 180 design. This agrees with the designed tip-tilt stability and falls within the expected jitter of SPHERE.
      \item Within expected manufacturing tolerances the coronagraph will be able to meet the requirements with a high degree of confidence ( more than $3 \sigma$).
      \item The most important aspect of the FIU is the alignment of the fiber with respect to the microlens center according to the tolerance simulations. The tolerance requirement can be achieved by using photonic crystal fibers with large mode field diameters.
   \end{enumerate}
With the SCAR coronagraph we meet the requirements for the characterization of Proxima b with SPHERE+ and high-resolution spectroscopy. Due to the simplicity of the optical setup only minor modifications are necessary to accommodate the SCAR coronagraph. Furthermore the single mode fibers simplify the design and decrease the size of the high-resolution integral field spectrograph as the input becomes diffraction limited. Therefore adding the new coronagraphic system as an upgrade to existing HCI instruments at current-generation telescopes will allow characterization of exoplanets in the solar neighborhood. An on-sky prototype is being built as the next step in the development process.

\begin{acknowledgements}
Haffert acknowledges funding from research program VICI 639.043.107, which is financed by The Netherlands Organisation for Scientific Research (NWO). Por acknowledges funding from NWO and the S\~{a}o Paulo Research Foundation (FAPESP). The research of Doelman and Snik leading to these results has received funding from the European Research Council under ERC Starting Grant agreement 678194 (FALCONER).
\end{acknowledgements}


\bibliographystyle{aa}
\bibliography{paper_references}

\end{document}